\documentclass[10pt,letterpaper]{article}
\usepackage{opex3}
\usepackage{color}
\usepackage{units}
\usepackage{amsmath}
\usepackage{amssymb}
\usepackage{graphicx}
\usepackage{bm}
\usepackage{multirow,color,relsize,ulem,microtype,mathtools}

\newcommand{\be}{\begin{equation}}
\newcommand{\ee}{\end{equation}}
\newcommand{\e}{\varepsilon}

\newcommand{\gper}{\gamma_\perp}
\newcommand{\gpar}{\gamma_\|}

\begin{document}

\title{Selective excitation of lasing modes by controlling modal interactions}

\author{Li Ge$^{1,2,*}$}

\address{$^1$Department of Engineering Science and Physics, College of Staten Island, CUNY, Staten Island, NY 10314, USA \\
$^2$The Graduate Center, CUNY, New York, NY 10016, USA}
\email{$^*$li.ge@csi.cuny.edu} 

\begin{abstract}
Using the correspondence between (saturated) nonlinear and (unsaturated) linear dielectric constants, we propose a simple and systematic method to achieve selective excitation of lasing modes that would have been dwarfed by more dominant ones of lower thresholds. The key element of this method is incorporating the control of modal interactions into the spatial pump profile, and it is most valuable in the presence of spatially and spectrally overlapping modes, where it would be difficult to achieve selective excitation otherwise.
\end{abstract}



The ability to selectively excite different modes of a given system not only reveals more information about the system itself but also enables a broad range of applications, such as magnetic resonance imaging \cite{Haacke}, coherent perfect absorption \cite{CPA,CPA2,CPA3}, and laser power enhancement \cite{Ge_NatPho}. Generally speaking, a target mode can be excited resonantly, using a monochromatic electromagnetic wave of the same frequency, if it is well separated spectrally from other modes in the system. This approach can be employed, for example, to excite a cavity mode that couples strongly to quantum emitters in cavity quantum electrodynamics \cite{Haroche,Devoret,Walther}. Also when combined with time-reversal symmetry and wavefront manipulation \cite{Fink}, even a typically strongly scattering system can be made scattering-free and a perfect absorber \cite{CPA,CPA2,CPA3}.

In optical systems where the energy is transferred indirectly via a material system (i.e., the ``gain medium"), a serious problem for selective mode excitation is that the excitation spectrum is often too broad to isolate a single optical mode.
One solution is to utilize the spatial intensity pattern of the target optical mode, i.e., by depositing focused energy (i.e., the ``pump") onto the target mode. This intuitive procedure, known as ``selective pumping," has been applied to both macro-cavity lasers \cite{Hermite,Hermite2,Woerdman} and micro-cavity lasers \cite{Microcavity1,Microcavity2,Microcavity3}, and it is a powerful tool to explore interesting phenomena such as exceptional points \cite{Liertzer} and chaos-assisted tunneling \cite{Narimanov}, besides reducing the threshold of a laser \cite{rex,spiral}, controlling its output directionality \cite{Rotter,Seng} and frequency \cite{Sebbah,Sebbah2,Fukushima}, and enhancing its output power \cite{Ge_NatPho,Woerdman}.

This intuitive approach, however, \textit{does not} work if the target mode has a relatively high loss (or equivalently, a relatively low quality factor) and strong spatial overlap with lower-loss modes. Simply focusing the pump onto the target mode still favors those lower-loss modes, whose thresholds are lower and whose intensities dwarf that of the target mode. Previous efforts based on numerical optimizations \cite{Seng,Sebbah,Sebbah2,Seng2} have shone some light on overcoming this hurdle, but they require many iterations of trial and error, and hence are computationally intense and do not suit applications such as optical switching.

In this work we propose a systematic method for selective mode excitation in lasers and other nonlinear optical media \cite{Deng,Snoke,Carusotto,GPE,Florent}, which addresses exactly this problem. The two key elements in our method are the following. First, we note that at any pump power $D_0$, the saturated nonlinear dielectric constant $\e(\vec{r};D_0)$ in the laser has an unsaturated linear correspondence $\tilde{\e}(\vec{r})$. They lead to the same set of lasing modes, with exactly the same lasing frequencies, spatial intensity patterns, but \textit{different} overall intensities. For example, we may find two lasing modes (1 and 2) at $D_0$ in a given laser cavity with a uniform spatial pump profile. Mode 1 has a lower threshold and its intensity is higher than mode 2. The saturated dielectric constant $\e(\vec{r};D_0)$ is different from its value at threshold, where gain saturation has just set in. Now by choosing a different pump profile, we can impose an unsaturated dielectric constant $\tilde{\e}(\vec{r})$ that matches exactly $\e(\vec{r};D_0)$. As we shall prove shortly, modes 1 and 2 are now at threshold simultaneously, while other modes are still below thresholds. In other words, by applying $\tilde{\e}(\vec{r})$ using the aforementioned correspondence between nonlinear and linear dielectric constants, we have lowered the threshold of mode 2 to be the same as mode 1, which is the first step in our proposal to selectively excite mode 2. Second, to suppress mode 1 while exciting mode 2, we modify the modal interactions manifested by $\e(\vec{r};D_0)$, by increasing the self-saturation of mode 1 or reducing that of mode 2. The resulting $\tilde{\e}(\vec{r})$ is no longer the same as $\e(\vec{r};D_0)$, and it can make the threshold of mode 2 considerably lower than mode 1, leading to a wide range of pump power in which mode 2 is the only lasing mode.

To further illustrate how this method works, below we discuss it in detail using the Steady-state Ab-initio Laser Theory (SALT) \cite{SALT,SPASALT,C-SALT}, which finds the steady-state solutions of the semiclassical laser equations \cite{Haken,Lamb}. We first briefly review SALT and use it to explain when the intuitive approach of selective pumping, i.e., focusing the pump onto the target mode, works and fails.

SALT assumes that the population inversion in the gain medium is stationary (see the discussion in Ref.~\cite{directMethod}), and the accuracy of SALT in this regime has been verified by comparing with time-dependent simulations \cite{OpEx08,OpEx11,OpEx12}.
In a steady state, the electric field is multi-periodic in time, i.e.,
\be
E^+(\vec{r},t) = \sum_{\mu=1}^N \Psi_\mu(\vec{r})\,e^{-i\Omega_\mu t},
\ee
where $N$ is the number of lasing modes and only positive frequency components are shown. At a given pump power $D_0$, measured by the population inversion of the gain medium it creates, 
the nonlinear lasing modes $\Psi_\mu(\vec{r};D_0)$ and the laser frequencies $\Omega_\mu$ can be obtained by solving the following set of coupled Helmholtz equations \cite{SPASALT}
\be
\hspace{-7pt}\left[ \nabla^2 + [\e_c(\vec{r}) + \e_g(\vec{r};D_0)]\Omega_\mu^2 \right]\Psi_\mu(\vec{r};D_0) = 0,\;(\mu=1,\ldots,N) \label{eq:SALT1}
\ee
in which we have taken the speed of light in vacuum to be unity. $\Psi_{\mu}(\vec{r};D_0)$ here is dimensionless, measured in its natural units of $e_c = \hbar\sqrt{\gpar\gper}/2g$, where $\gpar$ and $\gper$ are the inversion and polarization relaxation rates and $g$ is the dipole matrix element between the energy levels of lasing transition.

The saturated nonlinear dielectric constant $\e(\vec{r};D_0)$ mentioned in the introduction is given by the sum of $\e_c(\vec{r})$ and $\e_g(\vec{r};D_0)$ in Eq.~(\ref{eq:SALT1}). $\e_c(\vec{r})$ is the ``passive'' part of the cavity dielectric function, given by $n_c^2(\vec{r})$ in terms of the cavity refractive index. $\e_g(\vec{r};D_0)$ captures the ``active" part of the dielectric function \cite{SPASALT}, i.e.,
\be
\hspace{-5mm}\mathclap{\e_g(\vec{r};D_0)\hspace{-2pt} =\hspace{-2pt} \frac{\gper}{\Omega_\mu \hspace{-2pt}-\hspace{-2pt} \omega_a \hspace{-1pt}+\hspace{-1pt} i\gper}\frac{D_0 f_0(\vec{r})}{1\hspace{-1pt}+\hspace{-1pt}\sum_{\nu=1}^N\Gamma_\nu|\Psi_\nu(\vec{r};D_0)|^2},} \label{eq:epsg}
\ee
which contains the nonlinear spatial hole burning interactions beyond the standard 3rd-order approximation \cite{OpEx08}.
$\omega_a$ here is the atomic transition frequency, $\Gamma_\nu \equiv \gper^2/[\gper^2 + (\Omega_\nu-\omega_a)^2]$ is the Lorentzian gain curve evaluated at lasing frequency $\Omega_\nu$, and $f_0(\vec{r})\geq0$ is the spatial pump profile, which is normalized by $\int_\text{cavity} f_0(\vec{r}) d\vec{r} = S$, where $S=\int_\text{cavity} d\vec{r}$ is the length (area) of the cavity in one (two) dimension(s).

To select a certain higher-loss mode $\mu$, we search for a pump profile $f_0(\vec{r})$ that makes its threshold the lowest among all possible lasing modes. Instead of comparing their actual thresholds $D^{(\mu)}_{0,\text{int}}$ that depend on the spatial hole burning interactions, it is more convenient to work with the noninteracting thresholds $D^{(\mu)}_{0}$, defined by
\be
\left[ \nabla^2 \hspace{-2pt}+\hspace{-2pt} \left(\hspace{-1pt}\e_c(\vec{r}) \hspace{-1pt}+\hspace{-1pt} \frac{\gper D^{(\mu)}_0f_0(\vec{r})}{\Omega_\mu \hspace{-1pt}-\hspace{-1pt} \omega_a \hspace{-1pt}+\hspace{-1pt} i\gper}\hspace{-1pt}\right)\hspace{-1pt}\Omega_\mu^2\hspace{-1pt} \right]\hspace{-2pt}\Psi_\mu(\vec{r};D^{(\mu)}_0) = 0. \label{eq:SALT_TH}
\ee
We note that $\Psi_\nu=0$ for all modes at the lowest threshold (i.e., $D_0=D_{0,int}^{(1)}$) and the modal interactions vanish. In this case Eq.~(\ref{eq:SALT1}) is identical to Eq.~(\ref{eq:SALT_TH}) and $D_{0,int}^{(1)}=D_{0}^{(1)}$. In other words, if $D^{(\mu)}_{0}$ is the lowest noninteracting threshold, then mode $\mu$ also has the lowest threshold when spatial hole burning interactions are considered. Therefore, we can judge whether the target mode has the lowest actual threshold by comparing all noninteracting thresholds $D^{(\nu)}_{0}$.

Unless a mode is very lossy, the reduction of its threshold by selective pumping is given approximately by the pump overlapping factor \cite{Ge_NatPho}
\be
r_\mu = \frac{\int_\text{cavity} f_0(\vec{r})|\Psi_\mu(\vec{r};D_0)|^2 d\vec{r}}{\int_\text{cavity} |\Psi_\mu(\vec{r};D_0)|^2 d\vec{r}},\label{eq:fmu}
\ee
which becomes 1 for uniform pumping by definition (i.e., $r_\mu=1$ for $f_0(\vec{r})=1$).
Suppose that there are two modes (1 and 2) with distinct spatial profiles and that mode 2 has a higher threshold with uniform pumping. By focusing the pump spatially onto mode 2 (for example, with $f_0(\vec{r})\propto|\Psi_2(\vec{r};D_0)|^2$), $r_2$ can become much larger than 1 while $r_1$ unavoidably becomes much less than 1 (due to its distinct spatial profile from mode 2 and $f_0(\vec{r})$), which then makes $D^{(2)}_0<D^{(1)}_0$ and inverts the order of these two lasing modes, leading to the selective excitation of the higher-loss mode 2. When modes 1 and 2 overlap strongly in space however, one finds that $r_2\sim r_1$ when focusing the pump onto mode 2, meaning that the thresholds of modes 1 and 2 are reduced by a similar factor, and mode 1 remains the mode with the lowest threshold. In this case, the intuitive approach to selective excitation fails.

Having explained when focusing the pump onto the target mode works and fails to achieve selective excitation, below we show how the two-step approach outlined in the introduction works for our benefit. We first note that at any pump power $D_0$ above threshold, the nonlinear equation (\ref{eq:SALT1}) and the linear equation (\ref{eq:SALT_TH}) are no longer identical due to the non-zero spatial hole burning interactions in the former. However, with a new pump profile
\be
\tilde{f}_0(\vec{r})=\frac{Cf_0(\vec{r})}{1+\sum_{\nu=1}^N\Gamma_\nu|\Psi_\nu(\vec{r};D_0)|^2},\label{eq:Transformation}
\ee
the unsaturated linear dielectric constant $\tilde{\e}(\vec{r})=\e_c(\vec{r}) +(\gper \tilde{D}^{(\mu)}_0\tilde{f}_0(\vec{r}))/(\Omega_\mu -\omega_a + i\gper)$ in Eq.~(\ref{eq:SALT_TH}) becomes the same as the saturated nonlinear dielectric constant $\e(\vec{r};D_0)=\e_c(\vec{r};D_0)+\e_g(\vec{r};D_0)$ in Eq.~(\ref{eq:SALT1}) with the original pump profile $f_0(\vec{r})$, where
\be
\tilde{D}_0^{(\mu)}=\frac{D_0}{C} \label{eq:TH_transformed}
\ee
and $C$ is a normalization constant such that $\int_\text{cavity} \tilde{f}_0(\vec{r}) d\vec{r} = S$. Since Eq.~(\ref{eq:SALT1}) holds for all lasing modes at $D_0$, the correspondence described above implies that with the new pump profile $\tilde{f}_0(\vec{r})$, these modes have the \textit{same} noninteracting threshold given by Eq.~(\ref{eq:TH_transformed}), and they are the lowest among all $\tilde{D}_0^{(\nu)}$.
This is the first step in our method, which levels up the threshold of the target mode with all the lower-loss modes, and hence eliminates its disadvantage due to its higher loss.



In the second step, we modify $\tilde{f}_0(\vec{r})$ given by Eq.~(\ref{eq:Transformation}) such that it favors the target mode $\mu$. We will refer to the resulting pump profile as $\tilde{f}_\mu(\vec{r})$, and it can be chosen, as mentioned in the introduction, by increasing the overall intensity of $|\Psi_{\nu\neq\mu}(\vec{r};D_0)|^2$ and hence the self saturation of the non-targeted modes in Eq.~(\ref{eq:Transformation}), which further suppresses these modes (``approach 1").
Another option is to reduce the overall intensity of $|\Psi_{\mu}(\vec{r};D_0)|^2$ and hence the self saturation of the target mode $\mu$ (``approach 2"). It can be even made negative as long as the pump profile $\tilde{f}_\mu(\vec{r})$ is still non-negative everywhere.
One may also combine approaches 1 and 2 when necessary.

\begin{figure}[b]
\centering
\includegraphics[width=0.75\linewidth]{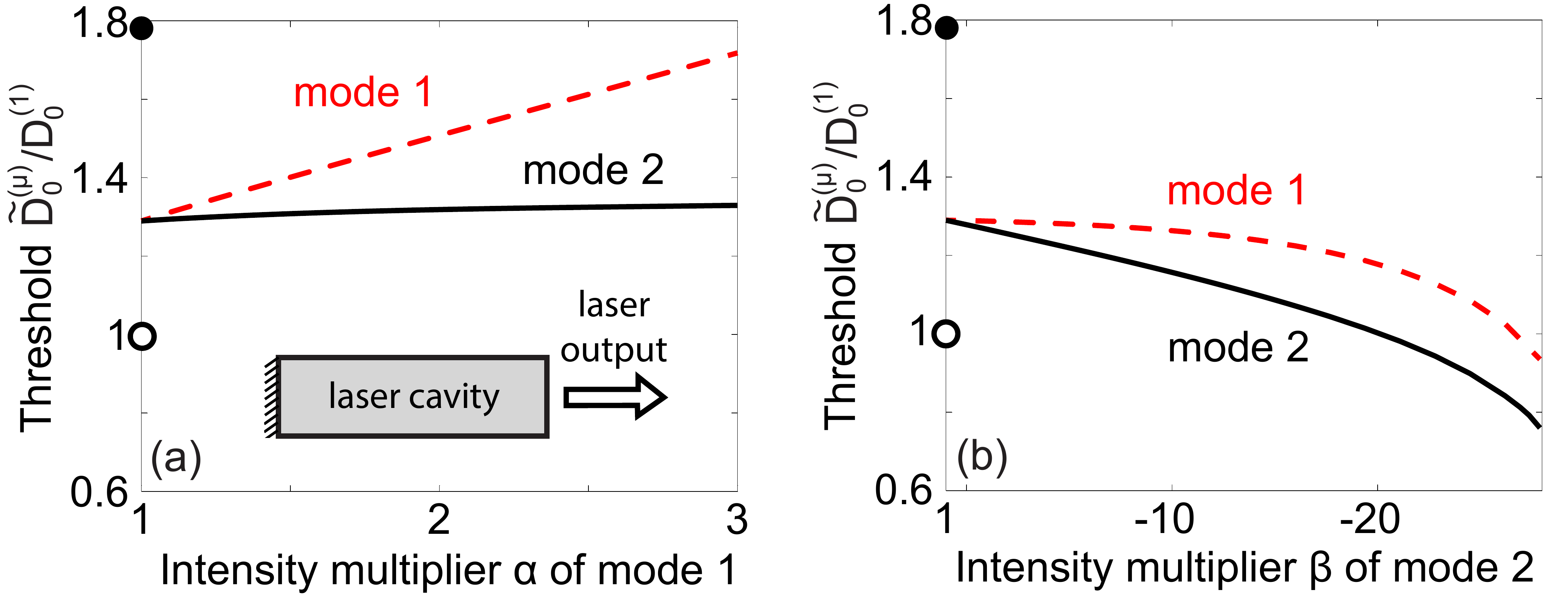}
\caption{(Color online) Thresholds in a 1D cavity with uniform pumping and selective pumping using the two-step method described in the main text. Open and filled dots show the actual thresholds of modes 1 and 2 with uniform pumping.
 Dashed and solid lines in both (a) and (b) show the noninteracting thresholds of mode 1 and mode 2, respectively. Their intersection on the vertical axis shows their identical threshold with the pump profile given by Eq.~(\ref{eq:Transformation}) at $D_0=1.88D_0^{(1)}$. In (a) we suppress mode 1 by multiplying its intensity in the original spatial hole burning interactions by a factor $\alpha\in[1,3]$. In (b) we favor mode 2 by multiplying its intensity (which is about 1/17 of that of mode 1) by a factor $\beta\in[-28,1]$.
Inset in (a): The cavity has refractive index $n_c=3$ and a perfect mirror on the left side. The gain medium is characterized by $\omega_aL=20$ and $\gper L=2$.
}\label{fig:1D}
\end{figure}
Below we exemplify the effectiveness of our method in a one-dimensional (1D) slab laser and a two-dimensional (2D) random laser. With uniform pumping, the slab laser of length $L$ shown in Fig.~\ref{fig:1D} first exhibits a lasing mode of frequency $\Omega_1L\simeq20.5$ at its threshold $D^{(1)}_0$, and we aim to selectively excite the second mode of frequency $\Omega_2L\simeq18.9$ and actual threshold $1.78D^{(1)}_0$.
If we focus the pump onto the second mode using $f_0(\vec{r})\propto|\Psi_2(\vec{r};D^{(2)}_0)|^2$, we find that the threshold of mode 2 is still 
higher than that of mode 1, even though it is reduced by 32\% from its value with uniform pumping. The result is much more promising when the pump profile is chosen according to the two-step method described above. By solving Eq.~(\ref{eq:SALT1}) at $D_0=1.88D^{(1)}_0$ with uniform pumping, we find the lasing modes $\Psi_1(\vec{r};D_0),\Psi_2(\vec{r};D_0)$ and subsequently $\tilde{f}_0(\vec{r})$ using Eq.~(\ref{eq:Transformation}). Indeed this $\tilde{f}_0(\vec{r})$ levels up the thresholds of modes 2 and 1 [$\tilde{D}^{(2)}_0=\tilde{D}^{(1)}_0=1.29D^{(1)}_0$; see the intersection of the solid and dashed lines in both panels of Fig.~\ref{fig:1D}]. We then modify this $\tilde{f}_0$ by gradually increasing the intensity of mode 1 [see Fig.~\ref{fig:1D}(a)] or decreasing the intensity of mode 2 [see Fig.~\ref{fig:1D}(b)]. Both approaches can create a considerable difference between $\tilde{D}_0^{(2)}$ and $\tilde{D}_0^{(1)}$, which is required for single-mode excitation of the target mode 2 in a wide range of pump power. We find that approach 2 is more favorable, since it leads to a threshold that is even lower than the lowest threshold with uniform pumping (i.e., ${D}^{(1)}_0$). 
\begin{figure}[b]
\centering
\includegraphics[width=0.75\linewidth]{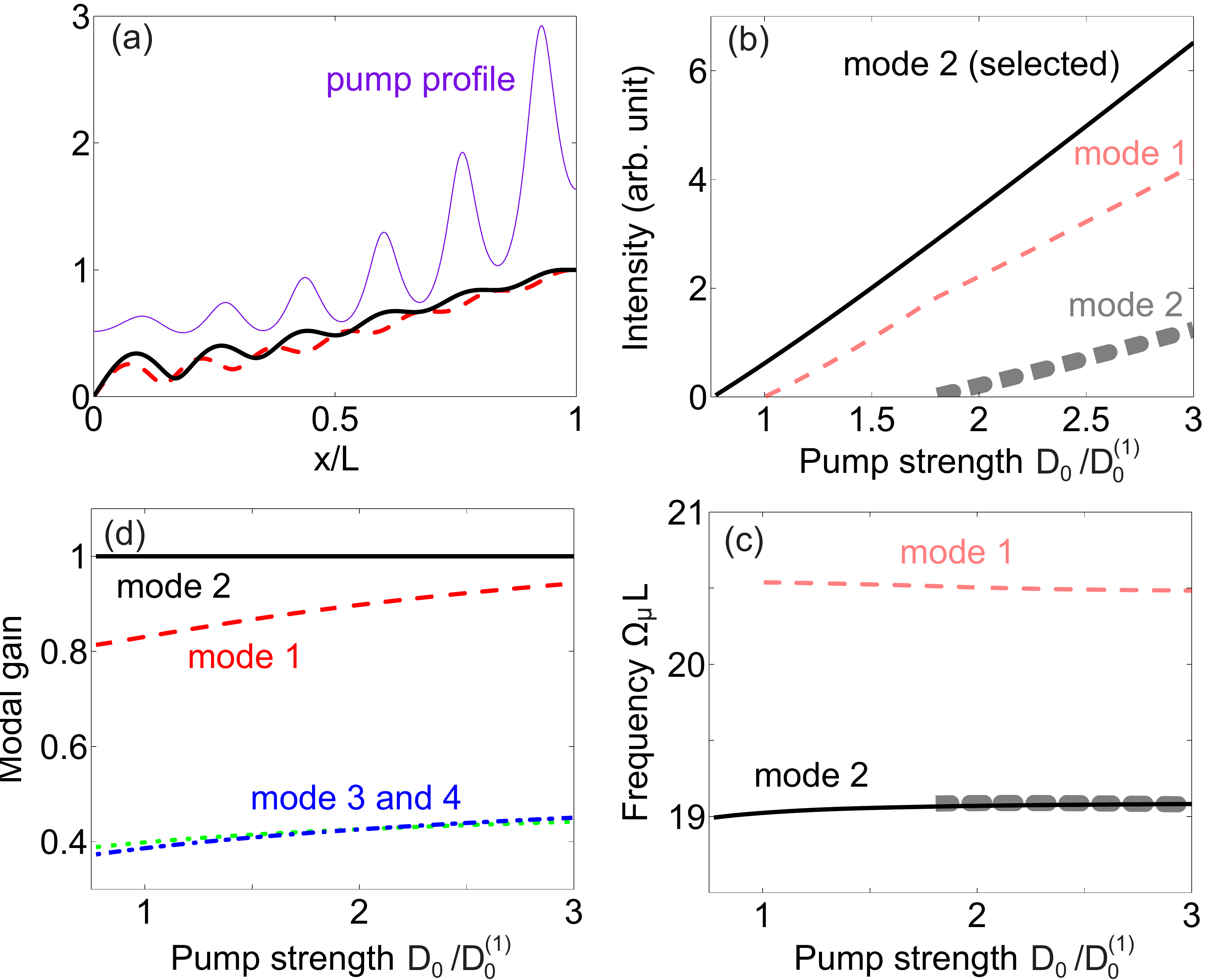}
\caption{(Color online) Reduced threshold and single-mode lasing using the two-step selective excitation described in the main text. (a) Pump profile $\tilde{f}_2(\vec{r})$ (purple thin solid line) that corresponds to the rightmost data in Fig.~\ref{fig:1D}(b). Also shown are the normalized mode profiles $|\Psi_1(\vec{r})|^2$ (red dashed line) and $|\Psi_2(\vec{r})|^2$ (thick black solid line) at $D_0=1.88D^{(0)}_1$ with uniform pumping. (b) Intensities at the right end of the cavity. With $\tilde{f}_2(\vec{r})$ in (a), the target mode 2 (black solid line) is the only lasing mode in the pump range shown. Red dashed line and black squares show the intensities of modes 1 and 2 with uniform pumping, respectively. The same legends are used in (c), which shows the frequencies of the lasing modes in (b). The left end of each line marks the threshold of the corresponding mode. (d) Modal gain of the first four modes with $\tilde{f}_2(\vec{r})$ in (a). }\label{fig:1D_NON}
\end{figure}

To confirm these observations which are based on the noninteracting thresholds of modes 1 and 2, we solve for the nonlinear lasing solutions with $\tilde{f}_2(\vec{r})$ that corresponds the rightmost data in Fig.~\ref{fig:1D}(b) [see Fig.~\ref{fig:1D_NON}(a)]. In comparison with uniform pumping, not only is the threshold of the target mode 2 reduced to 0.77$D_0^{(1)}$ with this $\tilde{f}_2(\vec{r})$, mode 2 is also the only lasing mode in the whole pump range shown in Fig.~\ref{fig:1D_NON}(b). The latter observation can be confirmed by calculating the modal gain \cite{SPASALT}: a mode becomes lasing if its modal gain reaches 1 from below, which then stays at 1 unless the mode is killed \cite{Liertzer,SALT}.
Indeed all the non-targeted modes have a modal gain below 1 in this pump range, as shown in Fig.~\ref{fig:1D_NON}(d). We also note that the intensity of mode 2 with $\tilde{f}_2(\vec{r})$ has a steeper slope than both modes 1 and 2 with uniform pumping [see Fig.~\ref{fig:1D_NON}(b)], indicating an improved utilization of the pump energy.

\begin{figure}[b]
\centering
\includegraphics[width=0.75\linewidth]{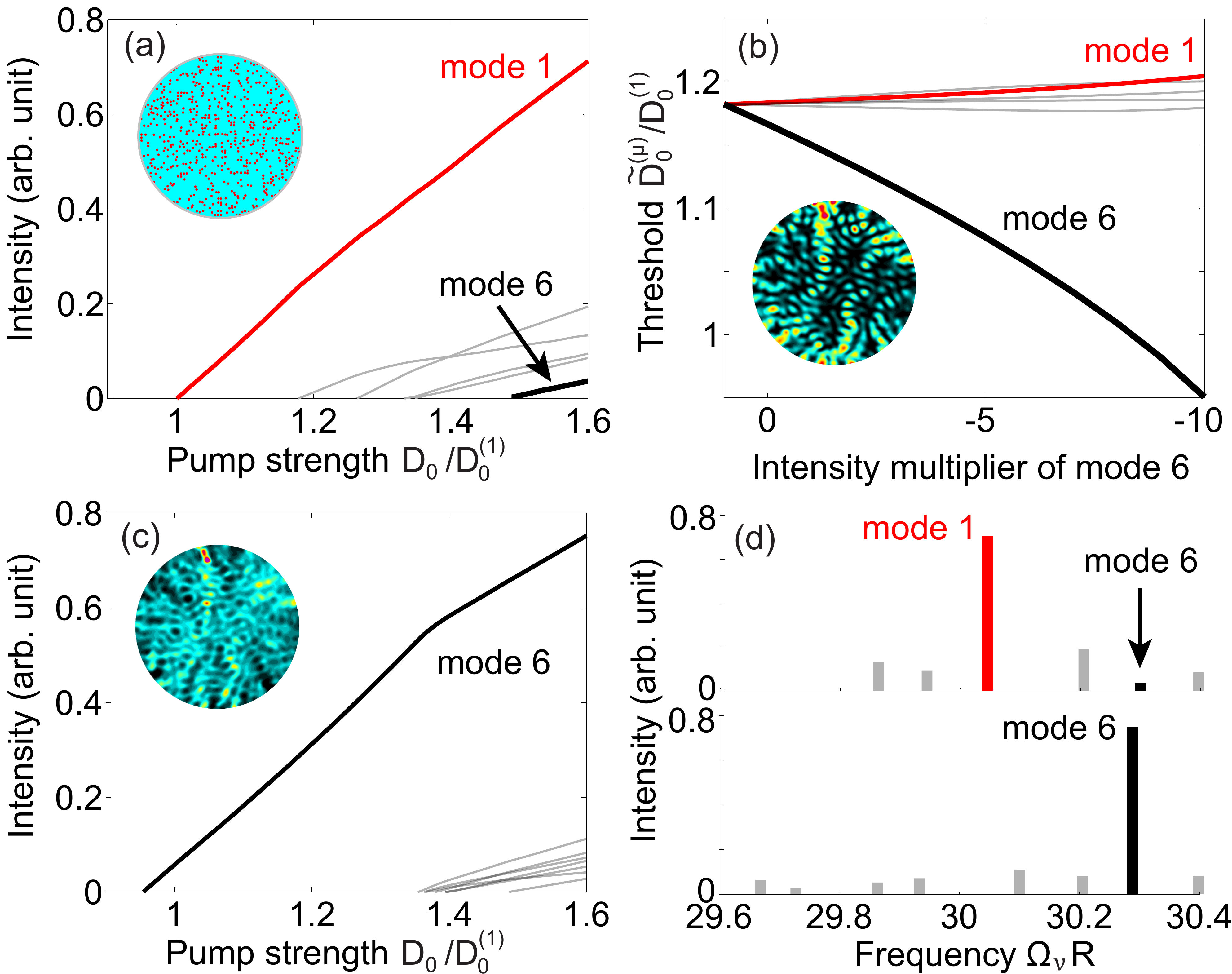}
\caption{(Color online) Selective excitation in a 2D diffusive random laser. (a) Intracavity intensity for the first six modes with uniform pumping. The black line shows the 6th mode to be selected. Inset: The system is modeled as a disk region of radius $R$ containing random scatterers of refractive index $n=1.2$ and a background index $n=1$. The gain medium is characterized by $\omega_aR=30$ and $\gper R=2$. (b) Noninteracting thresholds of the six modes in (a) but with a new pump profile given by Eq.~(\ref{eq:Transformation}) (the leftmost data points) and then gradually decreasing the intensity of the target mode 6 in the spatial hole burning interactions. Inset: False-color intensity plot of mode 6.
(c) Same as (a) but with the pump profile $\tilde{f}_6(\vec{r})$ shown in the inset [the rightmost data points in (b)].
(d) Spectra at $D_0=1.6D^{(1)}_0$ with uniform pumping (upper panel) and with $\tilde{f}_6(\vec{r})$ in (c) (lower panel).}\label{fig:2D}
\end{figure}

Having exemplified our method of selective excitation in the simple 1D slab laser, next we tackle a more complicated laser, a 2D diffusive random laser \cite{SALT,RLreview}, in which the lasing modes are strongly overlapping in space.
In the example shown in Fig.~\ref{fig:2D} there are six modes lasing at $D_0=1.6D_0^{(1)}$. As a challenge to our method, we target the 6th mode with the highest threshold and lowest intensity.
We carry out the nonlinear-linear correspondence using Eq.~(\ref{eq:Transformation}) at this pump power, after which all the six lasing modes have the same threshold [see the intersection point on the vertical axis in Fig.~\ref{fig:2D}(b)]. Next we follow approach 2 when modifying $\tilde{f}_0(\vec{r})$, by decreasing the intensity of mode 6 in the spatial hole burning interactions to $-10$ times. As a result,
the threshold of the target mode 6 is reduced to below $D^{(1)}_0$ and significantly lower than the other five modes.
If we choose $\tilde{f}_6(\vec{r})$ that corresponds to the rightmost data points in Fig.~\ref{fig:2D}(b), the target mode 6 becomes the only lasing mode until the pump power is 35\% above $D_0^{(1)}$ [Fig.~\ref{fig:2D}(c)], with slightly shifted frequency and more than ten-fold power increase [Fig.~\ref{fig:2D}(d)]. We note that mode 1 is suppressed in the pump range shown.

The same procedure has been applied to select modes 2 to 5, one at a time. In each case a significant pump range of single-mode operation is found for the target mode. These results highlight the generality of selective excitation based on the correspondence between (saturated) nonlinear and (unsaturated) linear dielectric constants, which applies to all nonlinear optical media in their steady states, including but not limited to lasers and exciton-polariton condensates \cite{Deng,Snoke,Carusotto,GPE,Florent}. We note that the first step in our method can be viewed as a special case of the second step, in which not only the self saturation of the non-targeted modes but also that of the target mode are increased from zero. Although the latter does not seem ideal and can be reversed in step 2, the very fact that different thresholds of all lasing modes with the original pump profile $f_0(\vec{r})$ level up after the first step is already a confirmation of the effectiveness of our method, the key element of which is incorporating the control of modal interactions into the pump profile. For experimental realizations of our proposal, the pump profile can be shaped via a spatial light modulator \cite{Seng,Sebbah, Sebbah2,SLM} for optically pumping and a pixelated contact for electrically pumping. We thank Hui Cao, Douglas Stone, Seng Fatt Liew, and Hakan T\"ureci for helpful discussions. This project is supported by PSC-CUNY 46 Research Grant from the City University of New York and NSF under grant No. DMR-1506987.

\end{document}